\documentclass[aps]{revtex4}
\usepackage{amssymb}
\usepackage{graphics}  

\usepackage{amsfonts}
\usepackage{amsmath}
\usepackage{amssymb}
\usepackage{graphicx}

\newcommand{\marge}[1]{\marginpar{}}  
\newcommand{\Sl}[1]{{}}           
\newcommand{\beq}[1]{\Sl{#1}\begin{equation}\if#1\empty\else\label{#1}\fi}
\newcommand{\eeq}{\end{equation}}
\newcommand{\beqa}[1]{\Sl{#1}\begin{eqnarray}\if#1\empty\else\label{#1}\fi}
\newcommand{\eeqa}{\end{eqnarray}}

\newcommand{\nm}{\nonumber\\}
\newcommand{\Eq}[1]{(\ref{#1})}

\usepackage{color}
\definecolor{red}{rgb}{1,0,0}

\providecommand{\U}[1]{\protect\rule{.1in}{.1in}}
\begin{document}
\title{Temporal Fokker-Planck Equations}
\author{Jean Pierre Boon}
\email{jpboon@ulb.ac.be}
\homepage{http://poseidon.ulb.ac.be}
\author{James F. Lutsko}
\affiliation{Centre for Nonlinear Phenomena and Complex Systems, CP 231, Universit\'e Libre de Bruxelles, 1050 - Bruxelles, Belgium}
\email{jlutsko@ulb.ac.be}
\homepage{http://www.lutsko.com}


\date{\today}
\maketitle

{\bf Abstract}

The temporal Fokker-Plank equation [{\it J. Stat. Phys.}, {\bf 3/4}, 527 (2003)] 
or propagation-dispersion equation 
was derived to describe diffusive processes with temporal dispersion
rather than spatial dispersion as in classical diffusion. 
We present two generalizations of the  temporal Fokker-Plank equation for 
the first passage distribution function $f_j(r,t)$ of a particle moving on a substrate 
with time delays $\tau_j$. Both generalizations follow from the first visit master
equation. In the first case, the time delays depend on the local concentration, that is
the time delay probability $P_j$ is a functional of the 
particle  distribution function and we show that when the functional dependence is of 
the power law type, $P_j \propto f_j^{\nu - 1}$, the generalized Fokker-Plank equation
exhibits a structure similar to that of the nonlinear spatial diffusion equation where 
the roles of space and time are reversed. In the second case, we consider the situation 
where the time delays are distributed according to a power law, 
$P_j \propto \tau_j^{-1-\alpha}$ (with $0 < \alpha < 2$), in which case we obtain 
a fractional propagation-dispersion equation which is the temporal analog of the fractional  
spatial diffusion equation (with space and  time interchanged).



\bigskip

\section{Introduction}
\label{intro}

Typical spatial diffusion processes are formulated in the continuum limit by the 
convection-diffusion equation whose solution is a Gaussian centered at
the most-likely position of a particle (a walker) moving at a constant velocity. Reciprocally 
there are situations in which, instead of asking where the walker would be after a given
time (long with respect to the duration of an elementary time step), one addresses the
the question as to how long it takes to reach a given point, at some large distance from the 
starting position (large compared to the unit length covered during the elementary time step).
For a stochastic process, one then asks what is the distribution of times taken to reach that 
point and the problem can be described by a propagation-dispersion equation giving a 
Gaussian time distribution centered at the most likely time of arrival at the target point 
\cite{boon-grosfils-lutsko}. This characterizes classical  time dispersion when the distribution  
originates from an Einstein type master equation \cite{einstein} for the probability 
$f(r,t)$ of finding the particle at position $r$ at time $t$ as briefly described in Sec.\ref{GenME}.
A practical example is given in \cite{hulin} which describes an experiment
where small beads are dropped into a container filled with larger beads. The small beads,
driven by gravity, diffuse through the array of larger beads and their collisions
with the larger ones induce time delays in the downward motion. Measurement of the
arrival times of the small beads at the end point of the container gives a Gaussian distribution,
(see Fig.9 in \cite{hulin}) i.e. the signature of a temporal-dispersion process. 

In many problems in physics, chemistry and biology, processes are time delayed or accelerated 
because they exhibit a functional dependence on the local concentration or on the time delays,
in which cases one expects deviations from the classical Gaussian distribution. Here we start
with the generalized master equation where the waiting time probability is a functional of the
distribution function $f(r,t)$ and in Sec.\ref{NonLinFPE} we derive a generalized Fokker-Plank 
equation (GFPE) for the first passage distribution function $f(r,t)$. 
Using a scaling argument (Sec.\ref{ScalingSec}) we obtain its solution
which is shown to exhibit a narrowing of the temporal distribution, i.e. temporal localization. 
Alternatively in Sec.\ref{fractional}   we introduce a power law {\it ansatz} for the time delay probability 
and we obtain a description of the evolution of the time distribution in the form of a fractional
temporal Fokker-Plank equation (FFPE). So it follows that the macroscopic evolution of the system 
is given by two complementary descriptions,  the nonlinear temporal Fokker-Plank equation 
or the fractional temporal Fokker-Plank equation, depending on the basic mechanisms of the 
time delay processes.

\section{Generalized master equation}
\label{GenME}

Consider a walker moving on a one-dimensional lattice whose sites are labeled by integers $l=0,\,1,\,2,\,...\,,\,n$.  The distance between neighboring sites  
is denoted by $\delta r$. The clock is set at $t_0=0$ when the particle is at site 
$l=0$ and its trajectory will intercept successively sites $l=1,\,2,\,3,\,...$ for the first time 
at times $t_1,\,t_2,\,t_3\,...$ . The $t_j$'s are integer multiples 
of the time step $\delta t$. While sites $0,\,1,\,2,\,3,\,...$ 
are equally spaced, the time differences between first visits, 
$t_{j+1} - t_j$, are (in general) not equally distributed.
Let $j_r$ be the random variable which corresponds to the number of steps 
required for the particle to reach position  $r = l_r\delta r$ for the first time at time  $t_r=j_r\delta t$.
We define $f(r,t)$ as the probability of finding the particle at position $r$ for the first 
time at time $t$;  $f(r,t)$ obeys the finite difference equation 
\beq{a0}
f(r,t)=\sum_{j=1}^N\,p_j\,f(r - \delta r,t-\tau_j)\;,
\eeq 
where $\tau_j = j\,\delta t$ and $p_j$ is the time delay probability, i.e. the probability that 
it takes $j$ time steps for the particle to move from site $r - \delta r$ to site $r$.
Equation (\ref{a0}) is  the {\em first visit equation} 
\cite{boon-grosfils-lutsko} which is the analog of Einstein's master equation for the 
classical random walk wherefrom the usual diffusion equation follows \cite{einstein}. 
In the hydrodynamic limit the first visit equation (\ref{a0}) yields  the
{\em propagation-dispersion equation} \cite{boon-grosfils-lutsko}
\begin{eqnarray}
\frac{\partial}{\partial r}f(r,t)\,+\,\frac{1}{c}\,\frac{\partial}
{\partial t} f(r,t)\,=\,
\frac{1}{2}\,{\gamma}\,\frac{\partial^2}{\partial {t^2}} f(r,t)\;,
\label{a1}
\end{eqnarray}
where the value of  $c^{-1}$ is given by the first moment $J_1 = \sum_{j=1}^N\,j\,p_j$,
and that of $\gamma$ by the second cumulant $\sum_{j=1}^N\,j^2\,p_j\,-\,J_1^2\,=\,J_2\,-\,J_1^2$;
$c$ is the propagation speed of the particle, and $\gamma$ the 
{\em time dispersion coefficient}.
Equation (\ref{a1}) is the analog of the advection-diffusion equation, but
describes a dispersion process in {\em time} (instead of diffusion
in space) with a drift expressed by a propagation speed with non-zero 
bounded values. The solution to Eq.\Eq{a1} is a Gaussian in time as illustrated in Fig.1.


Consider now that the waiting times depend on the particle distribution function
(in the example of the marathon (Fig.1) this corresponds to the 
local concentration of runners)  that is the  time delay probability $p_j$ in Eq.(\ref{a0}) 
is replaced by  a functional of $f(r,t)$
\beq{a3}
P_j \equiv p_j F_j^{(\nu)} [f]  
\eeq
with the normalization $\sum_j \, P_j \,=\sum_{j=1}^{N}p_{j}F_{j}^{(\nu)}\left[ f\right] = \, 1$, 
and where the index 
$\nu$ is such that  $F^{(\nu = 1)} [f]\,=\,1$. The functional plays the
role of a weighting factor to the amplitude of the waiting time probabilities. 
 $F_{j}^{(\nu)}\left[ f\right] $ is a functional
which means that, in principle, it depends on $f\left( r,t\right) ,f\left(
r,t-\delta t\right) ,f\left( r,t-2\delta t\right) $, ... For example, if we consider an 
algebraic function of the form 
$F_{j}^{(\nu)}\left[ f\right] $ $\sim f^{\nu - 1}\left( r,t-j\delta t\right) $,
then the normalization demands%
\begin{equation}
F_{j}^{(\nu)}\left[ f\right] =\frac{f^{\nu - 1}\left( r,t-j\delta t\right) }{%
\sum_{l=1}^{N}p_{l}f^{\nu - 1}\left( r,t-l\delta t\right) }\,.
\label{nu_norm}
\end{equation}
So for clarity, we should  write explicitly
\begin{equation}
F_{j}^{(\nu)}\left[ f\right] =F_{j}\left[ f\left( r,t-j\delta t\right) ;f\left(
r,t-\delta t\right) ,f\left( r,t-2\delta t\right) ,...,f\left( r,t-N\delta t\right) \right] \,,
\label{Fnu}
\end{equation}
which allows for an explicit dependence on the index $j$. 
A slightly more restricted form which does not include the $j$
dependence, i.e. $F_{j}\left[ f_j; f_{1},...,f_{N}\right] =F\left[f_j; f_{1},...,f_{N}\right] $ 
will be considered below.
With (\ref{a3}), Eq.(\ref{a0}) becomes the {\it generalised master equation}
\beq{a4}
f(r + \delta r , t) - f(r, t) =
\sum_{j = 1}^n\,p_j F_j^{(\nu)} [f_j] \,\left(f(r , t- j \delta t) - f(r, t)\right)\;.
\eeq

\section{\protect\bigskip Nonlinear Fokker-Planck equation}
\label{NonLinFPE}

We consider the expansion of $F_{j}^{(\nu)}\left[ f\right] $ (for simplicity 
in the notation we shall omit the upper index ${(\nu)}$ which will be reintroduced 
when necessary):%
\begin{align}
F_{j}\left[ f\right] & =\left\{ F_{j}\left( x;y_{1}...,y_{N}\right) \right\}
_{f} 
 -\delta t\left\{ j\frac{\partial F_{j}\left( x;y_{1}...,y_{N}\right) }{%
\partial x}+\sum_{l=1}^{N}l\frac{\partial F_{j}\left(
x;y_{1}...,y_{N}\right) }{\partial y_{l}}\right\} _{f}\left(\frac{\partial f\left(
r,t\right) }{\partial t} \right) +...  \,, 
\label{norm_cond}
\end{align}
where the notation $\left\{ ...\right\} _{f}$ means that all the variables $%
x,y_{1},...$ are to be set equal to the $f\left( r,t\right) $'s as on the r.h.s. of Eq.\Eq{Fnu}.
Using this expansion, the generalized master equation \Eq{a4} becomes%
\begin{align}
&\delta r\frac{\partial f\left( r,t\right) }{\partial r}+\frac{1}{2}\left(
\delta r\right) ^{2}\frac{\partial^{2}f\left( r,t\right) }{\partial r^{2}}%
+... =-\delta t\sum_{j=1}^{N}jp_{j}\left\{ F_{j}\left(
x;y_{1}...,y_{N}\right) \right\} _{f}\frac{\partial f\left( r,t\right) }{%
\partial t} \notag\\
& +\frac{1}{2}\left( \delta t\right) ^{2}\sum_{j=1}^{N}jp_{j}\left\{
F_{j}\left( x;y_{1}...,y_{N}\right) \right\} _{f}\frac{\partial^{2}f\left(
r,t\right) }{\partial t^{2}} \notag \\
& +\left( \delta t\right) ^{2}\sum_{j=1}^{N}jp_{j}\left\{ j\frac{\partial
F_{j}\left( x;y_{1}...,y_{N}\right) }{\partial x}+\sum_{l=1}^{N}l\frac{%
\partial F_{j}\left( x;y_{1}...,y_{N}\right) }{\partial y_{l}}\right\}
_{f}\left( \frac{\partial f\left( r,t\right) }{\partial t}\right) ^{2} 
+...  \,.
\end{align}

By multiscale expansion and using the normalization condition 
(see details in Appendix \ref{appendixA}) we obtain
\beqa{a7}
\frac{\partial f \left( r,t\right) }{\partial r}&+&%
J_1\left[f \right]\,\frac{\delta t}{\delta r}\,
\frac{\partial f \left( r,t\right) }{\partial t} \nm%
\,&=&\,\left( J_2 \left[f \right]\,-\,\left(J_1\left[f \right]\right)^2 \right)\,\frac{(\delta t)^2}{2 \,\delta r}\,\frac{\partial^2 f (r,t)}{\partial t^2}\,+\, {\Lambda }\left[f \right] \frac{(\delta t)^2}{\delta r}
\left(\frac{\partial f(r,t)}{\partial t}\right)^2 \,,
\label{GFPE1}
\eeqa
where the $J_i$'s are the generalized moments
\begin{align}
&J_1\left[f \right]\,=\,
\sum_{j=1}^{N}jp_{j}\left\{ F_{j}\left( x;y_{1}...,y_{N}\right)\right\} _{f} \,,  \\
&J_2 \left[f \right]\,=\,\sum_{j=1}^{N}j^{2}p_{j}\left\{ F_{j}\left( x;y_{1}...,y_{N}\right) \right\}
_{f} \,, \\
&\Lambda\left[f \right]\,=\,\sum_{j=1}^{N}jp_{j}\left\{ j\frac{\partial F_{j}\left(
x;y_{1}...,y_{N}\right) }{\partial x}+\sum_{l=1}^{N}l\frac{\partial
F_{j}\left( x;y_{1}...,y_{N}\right) }{\partial y_{l}}\right\} _{f} \notag \\ 
&\;\;\;\;\;\;\;\;\;\;\;\;-\sum_{j=1}^{N}jp_{j}\left\{ F_{j}\left( x;y_{1}...,y_{N}\right) \right\}_{f} \times \notag \\
&\;\;\;\;\;\;\;\;\;\;\;\;\times \sum_{l=1}^{N}lp_{l}\left\{ \frac{\partial F_{l}\left(
x;y_{1}...,y_{N}\right) }{\partial x}+\sum_{k=1}^{N}%
\frac{\partial F_{l}\left( x;y_{1}...,y_{N}\right) }{\partial y_{k}}\right\}_{f}\,.%
\label{Js}
\end{align}
Equation \Eq{GFPE1} gives the general form of the 
{\em generalized temporal Fokker-Planck equation} (GFPE). 

We consider the case where $F_{j}\left( x;y_{1}...,y_{N}\right)$ does not depend 
explicitly on $j$, i.e.  $F_{j}\left( x;y_{1}...,y_{N}\right)=F\left( x;y_{1}...,y_{N}\right)$, 
in which case the normalization conditions \Eq{norm1}  imply 
\begin{equation}
0=\left\{ \frac{\partial F\left( x;y_{1}...,y_{N}\right) }{\partial x} %
+\sum_{l=1}^{N}\frac{\partial F\left( x;y_{1}...,y_{N}\right) }{\partial
y_{l}}\right\} _{f} \,,
\label{norm3}
\end{equation}%
so that the second term on the r.h.s. of $\Lambda\left[f \right]$ in \Eq{Js} vanishes, 
and we have
\beqa{J1_J2}
J_1[f] &=& J_1= \sum_j j\,p_j \;\;\;;\;\;\;J_2 [f]\,=\,J_2 = \sum_j\,j^2 p_j  \,,  \notag  \\
\Lambda \left[f \right]\,&=&\,\left\{ J_2\frac{\partial F_{j}\left(x;y_{1}...,y_{N}\right)}{\partial x}+\,%
J_1\sum_{l=1}^{N}\,l\,\frac{\partial F \left( x;y_{1}...,y_{N}\right) }{\partial y_{l}}\right\}_{f}  \nm
&=\,&\left( J_2 \,-\,J_1^2 \right)\,\left\{ \frac{\partial F\left( x;y_{1}...,y_{N}\right) }{%
\partial x}\right\} _{f}\,.
\label{Js2}
\eeqa
The  generalized Fokker-Planck equation \Eq{GFPE1} then becomes
\begin{equation}
\frac{\partial f\left( r,t\right) }{\partial r}+\frac{1}{c}%
\frac{\partial f\left( r,t\right) }{\partial t}=
\frac{1}{2}\,\gamma \left[ \frac{\partial ^{2}f\left( r,t\right) 
}{\partial t^{2}}+ 2 \left\{ \frac{\partial F^{(\nu)}\left( x;y_{1}...,y_{N}\right) }{%
\partial x}\right\} _{f}\left( \frac{\partial f\left( r,t\right) }{\partial t%
}\right) ^{2}\right]\,,
\label{GFPE2} 
\end{equation}
where 
\beqa{a13}
c^{-1}\,= \sum_j p_j \,j\,\frac{\delta t}{\delta r}\;,\;\;\;\;\;\;\;\;\;\;
\gamma\,=\,\left[ \sum_j p_j \,j^2 \,-\,\left(\sum_j p_j \,j\right)^2\right]\,\frac{(\delta t)^2}{\delta r}  
\eeqa
are the reciprocal propagation speed and the temporal dispersion coefficient respectively.
In Eq. \Eq{GFPE2}  we have reintroduced the index $\nu$ for later discussion.
Since by definition  $F^{(\nu = 1)} [f]\,=\,1$, it is clear that for $\nu = 1$, Eq.\Eq{GFPE2}
(as well as  \Eq{GFPE1}) reduces to the usual propagation-dispersion equation (\ref{a1}).

\section{Scaling and power law distribution}
\label{ScalingSec}

We now ask for a scaling solution of the GFPE 
\beq{scaling}
f\left( r,t\right) = r^{-\beta }\phi \left( \frac{t-r/c}{r^{\beta }}\right) \,,
\eeq
which by substitution in \Eq{GFPE1}  with $\zeta = \frac{t-r/c}{r^{\beta }}$  gives
\begin{equation}
-\phi \left( \zeta\right) -\zeta\phi ^{\prime }\left( \zeta\right)  = 
\beta ^{-1}r^{1-2\beta } \frac{1}{2}\,\left[ \left( J_2 \left[f \right]\,-\,\left(J_1\left[f \right]\right)^2 \right)\,\phi ^{\prime \prime}\left( \zeta\right) + 2\, r^{-\beta } \Lambda\left[f \right]\,
\phi ^{\prime 2}\left( \zeta\right) \right]   \,.
\label{eq_phi1}
\end{equation}%
The scaling equation can be satisfied  either with $\Lambda\left[f \right] = 0$
or with $\Lambda\left[f \right] \sim 1/f \left( r,t\right) $ and, in both cases $\beta =1/2$. 
The first case is realized for $\nu = 1$, i.e. $F^{(\nu)} = 1$; then 
$\left( J_2 \left[f \right]\,-\,\left(J_1\left[f \right]\right)^2 \right)\,{(\delta t)^2}/{\delta r} = \gamma$ and $J_1\left[f \right]\,{\delta t}/{\delta r} = c^{-1}$,
and Eq.\Eq{GFPE1} reduces to the classical temporal Fokker-Planck equation 
with Gaussian solution  \cite{boon-grosfils-lutsko}.
The second case can be satisfied if we require that $F^{(\nu)}{\left[f \right]}$ be a normalized 
power law independent of $j$. Indeed when  $F^{(\nu)}{\left[f \right]}$ does not dependent on $j$, 
$\Lambda\left[f \right]$  (\ref{Js2}) reduces to  
$\left( J_2 \,-\,\left(J_1\right)^2 \right)\,\left\{ \frac{\partial F^{(\nu)}
\left( x;y_{1}...,y_{N}\right) }{\partial x}\right\}_{f}$, 
and from the normalization condition, we have
\begin{equation}
F^{(\nu)}\left( x;y_{1}...,y_{N}\right) =\frac{K(x)}{p_{1}K(y_{1})+...+p_{N}K(y_{N})}\,,%
\end{equation}%
which gives%
\begin{align}
\left\{ \frac{\partial F^{(\nu)}\left(x;y_{1}...,y_{N}\right) }{\partial x}\right\}_{f}
=\frac{1}{p_{1}+...+p_{N}}\frac{1}{K(f)}\frac{\partial K(f)}{%
\partial f}  =\frac{\partial \ln K(f)}{\partial f} \,.
\end{align}%
The demand $ \Lambda\left[f \right] \sim 1/f \left( r,t\right)$ implies 
that, for some constant $\nu \geq 1$
\begin{equation}
\frac{\partial \ln K(f)}{%
\partial f}=\frac{\nu - 1}{f}  \;\;\;\;  \Longrightarrow \;\;\;\;\; K(f)=f^{\nu - 1} \,,
\end{equation}%
so that
\begin{align}
F^{(\nu)}\left(x;y_{1}...,y_{N}\right) & =\frac{x^{\nu - 1}}{p_{1}y_{1}^{\nu - 1}
+...+p_{N}y_{N}^{\nu - 1}} \,.
\label{power}
\end{align}%
Notice that the power law form follows from the scaling.
With this result and with $\beta = 1/2$, Eq.\Eq{eq_phi1}  becomes%
\begin{equation}
-\phi \left( \zeta\right) -\zeta\phi ^{\prime }\left( \zeta\right) =  \, \gamma \left[
\phi ^{\prime \prime }\left( \zeta\right) + 2(\nu - 1) \, \phi ^{-1}\left( \zeta\right) \phi
^{\prime 2}\left( \zeta\right) \right] \,.
\label{eq_phi2}
\end{equation} %
We also note that if one uses \Eq{power} as an {\em ansatz} in the GFPE (Eq. \Eq{GFPE2} )
 one obtains
\begin{equation}
\frac{\partial f\left( r,t\right) }{\partial r}=
\frac{1}{2}\,\gamma \left[ \frac{\partial ^{2}f\left( r,t\right) 
}{\partial t^{2}}+ 2\, \frac{\nu - 1}{f\left( r,t\right)} \left( \frac{\partial f\left( r,t\right) }{\partial t%
}\right) ^{2}\right]\,,
\label{GFPE3} 
\end{equation} 
which with the scaling relation \Eq{scaling}  gives exactly \Eq{eq_phi2}.

Equation \Eq{eq_phi2} can be simplified by introducing a change of variables and a transformation %
\begin{align}
x =\zeta/\sqrt \gamma =  \frac{t-r/c}{\sqrt{\gamma\,r}}\;\;\;\;;\;\;\;\;
\phi \left( \zeta\right)  =w^{\frac{1}{2\nu - 1 }}\left( x\right)  \,, 
\end{align}%
giving%
\begin{equation}
\frac{d^{2}w}{dx^{2}}+x\frac{dw}{dx}+\left( 2\nu - 1 \right) w=0 \,,
\label{eq_w}
\end{equation}%
which equation can be matched to the general confluent equation (see Appendix B)
and has the general solution
\begin{equation}
w\left( x\right) = A \exp \left( -\frac{x^{2}}{2} \right) M\left(1 - {\nu },\frac{1}{2},\frac{x^{2}}{2}\right) \,
+B  \,x\, \exp \left( -\frac{x^{2}}{2}\right) M\left( \frac{3}{2} - {\nu },\frac{3}{2},\frac{x^{2}}{2}\right) \,,
\label{gen_sol}
\end{equation}%
where $A$ and $B$ are constants and $M(a, b, x)$ is the confluent hypergeometric function.
Since the solution must be even in $x$, $B$ must be zero for symmetrical reasons. 
So the scaling distribution reads %
\begin{eqnarray}%
f\left( r,t\right) &= & \frac{1}{\sqrt {\gamma\,r}}\; \phi \left( \frac{t-r/c}{\sqrt {\gamma\,r}}\right) \, \notag \\
& = & \frac{1}{\sqrt {\gamma\,r}}\;\,\left[ A\,M\left(1- {\nu },\frac{1}{2},\frac{(t-r/c)^2}{2\,\gamma\,r}\right) %
\exp \left( -\frac{(t - r/c)^2}{2\,\gamma\,r}\right)  \right]^{\frac{1}{2\nu - 1}} \,.
\label{solution}
\end{eqnarray}%

For $\nu = 1$,  $M(0, \frac{1}{2}  ,\frac{x^{2}}{2}) = 1$; consequently   
in order to  retrieve the normalized Gaussian solution,  $A$ must be $\sqrt {2/\pi}$
and the final solution is given by
\begin{equation}%
f\left( r,t\right) =  
\frac{1}{\sqrt{\gamma\,r}}\,\left[ {\sqrt \frac{2}{\pi}}{M_{(1- \nu )}} \,%
\exp \left( -\frac{(t - r/c)^2}{2\,\gamma\,r} \right) \right]^{\frac{1}{2\nu - 1}} \,,
\label{solution2}
\end{equation}%
where ${M_{(1- {\nu )}}}= M(1-\nu,\frac{1}{2},\frac{x^{2}}{2})$ with $x=\frac{t - r/c}{\sqrt {\gamma\,r}}\,$\,.
Figure 1 illustrates this result for different values of the exponent. 

For $\nu = 3/2$,  we have
$ M(-  \frac{1}{2}, \frac{1}{2}  ,\frac{x^{2}}{2})
= \frac{1}{\sqrt 2}\,\exp(\frac{x^2}{4})\,E_1^{(0)}(x)$
(where $E_1^{(0)}(x)$ is the parabolic cylinder functions) 
giving
 \begin{equation}%
f_{\nu=3/2}\left( r,t\right) =  
\frac{1}{\sqrt{\gamma\,r}}\,%
\left( { {{\frac{1}{\sqrt \pi}}}}\; {E^{(0)}_{1}}\left( \frac{t - r/c}{\sqrt{\gamma\,r}} \right)\,%
\exp - \left( \frac{t - r/c}{2\,\sqrt{{\gamma\,r}}} \right)^{2}\,\right)^{1/2} \;.
\label{solution3/2}
\end{equation}%
For $\nu > 3/2$, we have $ M(-\frac{m}{2}, \frac{1}{2}  ,\frac{x^{2}}{2})$, %
but the  confluent hypergeometric function with $m>1$ 
exhibits alternating positive and negative regions separated by a singularity and consequently so for the distribution function; therefore values of $\nu>3/2$ must be physically rejected and
the meaningful range of the exponent is   $1\leq \nu \leq 3/2$, as illustrated in Fig.1. So when the nonlinear exponent $\nu$ increases we observe a narrowing of the distribution function that is a localisation in
temporal dispersion.

The asymptotic behaviour of the distribution follows from the observation that
for $|x|$ large (see  AS 13.5.1 in \cite{abramow})
\begin{equation}
M\left(a, b, x \right) = \frac{\Gamma(b)}{\Gamma(a)}\,e^x\,x^{a-b}\,\left(1+ {\cal O}(|x|^{-1}\right)\;\;\;\;;\;\;\;
x \in {\Re}, \, >0 \,,
\end{equation}%
which when used in (\ref{solution2}) gives 
\begin{equation}%
f\left( x >>1\right) 
\simeq \frac{1}{\sqrt {\gamma\,r}}\,\left(\frac{1}{\sqrt {\pi}}\,\frac{\Gamma\left(\frac{1}{2}\right)}%
{\Gamma\left(1-\nu \right)}\,\left(\frac{x^2}{2}\right)^{\frac{1-2\nu}{2}} 
\right)^{\frac{1}{2\nu - 1}} \notag \\
\label{solutionLargez}
\end{equation}%
or, with $t >> r/c$,
\begin{equation}
f\left( r, {t } >> {\sqrt {\gamma\,r}}  \right)\simeq \left({\frac{1}{\Gamma \left(1-\nu \right)}} %
\right)^{\frac{1}{2\nu - 1}}\,\frac{\sqrt 2}{t } \,,
\label{solutionLarge_t}
\end{equation}%
that is, for long times, $f\left( r, {t }  \right) \sim \,{t}^{-1}$, which is in reasonable agreement with the observation of time delays in earthquake distributions \cite{abe} 
$P(t) \sim t^{-\gamma}$ with ${\gamma} \simeq 0.9$.

\begin{figure}[tbp]
\begin{center}
\resizebox{\textwidth}{!}{
\includegraphics{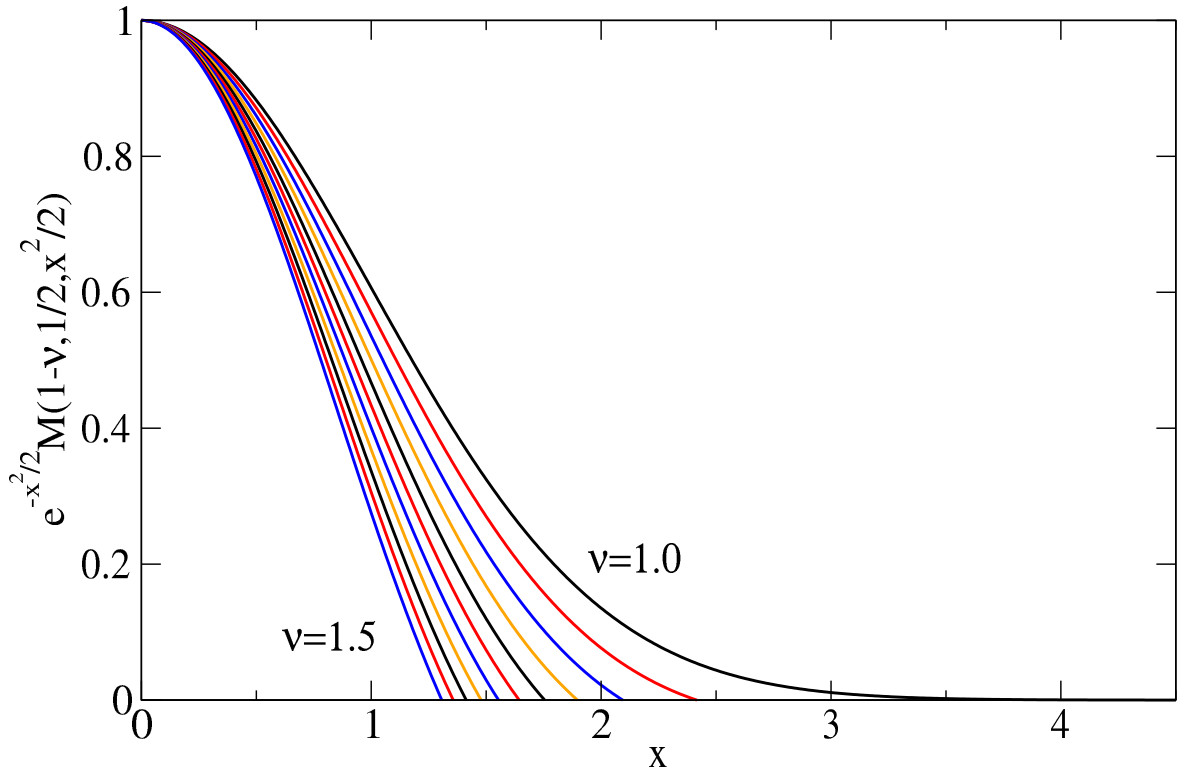}
}
\caption{ 
$M(1-\nu, \frac{1}{2}  ,\frac{x^{2}}{2})\,\exp \left(-\frac{x^{2}}{2}\right)$ (Eq.(28)) as a function of $x$
for various values of the exponent $1\leq\nu\leq3/2$.}
\end{center}
\end{figure}


\section{Fractional Fokker-Planck Equation}
\label{fractional}

The generalization of temporal diffusion to nonlinear jump probabilities discussed so far was developed based on a multiscale expansion that is only valid when the first and second moments of the jump probability exist. We now consider a second generalization that applies when the second moment does not exist. Unlike the nonlinear case, we will only consider processes for which the jump probabilities are statistically independent. When the second moment exists, this then leads via the central limit theorem to the classical Gaussian time distribution (see Fig.1) since for  the lattice model described in section \ref{GenME} 
the probability to reach lattice postition $l$ in time is simply the sum of the independent random waiting times $\widehat{t}=\sum_{l =1}^{n}\widehat{t}_{l}$. Explicitly, for large $l$, that is $ r >> \delta r$,
the probability for the stochastic variable $\widehat{t}$  to lie in the interval $[T,T+dT]$ is given by a Gaussian%
\begin{equation}
f(T; r)=\sqrt{\frac{1}{2\pi \sigma }}\exp \left( -\frac{%
\left( T-\overline{t}\left( r\right) \right) ^{2}}{2\sigma }\right) \;,
\label{2}
\end{equation}%
where the most likely time is%
\begin{equation}
\overline{t}=\sum_{l=1}^{n}\left\langle \widehat{t}_{l}\right\rangle
=n\int_{0}^{\infty }d\tau \; \tau p\; \left( \tau \right)  \label{3}
\end{equation}%
where $p\left( \tau \right)$ is the time delay probability, and the width of the distribution is%
\begin{equation}
\sigma (r)=\sum_{l=1}^{n}\left( \left\langle \widehat{t}_{l}^{2}\right%
\rangle -\left\langle \widehat{t}_{l}\right\rangle ^{2}\right)
= n \int_{0}^{\infty }d\tau \;\tau ^{2}p\left( \tau \right) -\frac{1}{n}\,%
\overline{t}^{2}.  \label{4}
\end{equation}

Consider now the case of distributions which do not possess second moments.
In particular, we will consider a power law distribution %
\begin{equation}
p(t) =  \frac{\alpha t_{0}^{\alpha }}{t^{\alpha +1}}\Theta(t-t_{0})\;\;\;;\;\;\;0<\alpha <2\;\;;\;\;\; t\rightarrow \infty  
\label{5}
\end{equation}%
and let $\widehat{t}=\frac{1}{N^{1/\alpha}}\sum_{l =1}^{n}\widehat{t}_{l}$.
The probability for $\hat{t} =T_{\alpha}$ is 
\begin{eqnarray}
f_{\alpha}(T_{\alpha},N) & = \int_{0}^{\infty} \delta(T_{\alpha}-\frac{1}{N^{1/\alpha}}\sum_{i=1}^{N}t_{i})p(t_{1})...p(t_{N})dt_{1}...dt_{N} \notag \\
\end{eqnarray} 
so that 
\begin{equation}
\tilde{f}_{\alpha}(\omega,N) \equiv \int_{-\infty}^{\infty} e^{i \omega T}f_{\alpha}(T_{\alpha}) = \left( \tilde{p}\left(\frac{\omega}{N^{1/\alpha}}\right)\right)^N
\end{equation} 
where a simple calculation gives
\begin{equation}
\tilde{p}(\omega) = \alpha (-i\omega t_{0})^{\alpha}\Gamma(-\alpha,-i \omega t_{0}) 
\end{equation}
which has the expansion for small $\omega t_{0}$ 
\begin{equation}
\tilde{p}(\omega) = \alpha (-i\omega t_{0})^{\alpha}\Gamma(-\alpha)-\alpha \sum_{k=0}^{\infty}\frac{(-i \omega t_{0})^{k}}{k!(k-\alpha)}
\end{equation}
Thus, taking the inverse Fourier transform gives
Expansion in $1/N$ leads to the result
\begin{eqnarray}
f_{\alpha}(T_{\alpha},N) & = & \int_{-\infty}^{\infty} \left( \tilde{p}\left(\frac{\omega}{N^{1/\alpha}}\right)\right)^N e^{-i \omega T_{\alpha}} \frac{d\omega}{2\pi} \notag \\
 & = & \int_{-\infty}^{\infty} exp\left( -i \omega T_{\alpha}+N \ln \tilde{p}\left(\frac{\omega}{N^{1/\alpha}}\right) \right) \frac{d\omega}{2\pi} \notag \\
& = & \int_{-\infty}^{\infty} exp\left (-i \omega T_{\alpha} +\alpha \left (-i \omega t_{0}\right)^{\alpha} \Gamma(-\alpha) + \frac{\alpha}{1-\alpha} (-i \omega t_{0})N^{1-1/\alpha}  +O(N^{1-2/\alpha})\right ) \frac{d\omega}{2\pi} \notag \\
\end{eqnarray}
so that the higher order terms are negligable for large $N$ provided $\alpha < 2$. The probability density that the first arrival time to reach position $N$ is $T = N^{1/\alpha}T_{\alpha}$ is therefore
\begin{eqnarray}
f(T,N) & = & \int_{0}^{\infty} f_{\alpha}(T_{\alpha},N)\delta(T-N^{1/\alpha}T_{\alpha}) dT_{\alpha} \notag \\
& = & N^{-1/\alpha}\int_{-\infty}^{\infty} exp\left (-i \omega N^{1-1/\alpha} \left(\frac{T}{N}  + \frac{\alpha}{1-\alpha}  t_{0}\right) +\alpha \left (-i \omega t_{0}\right)^{\alpha} \Gamma(-\alpha) +O(N^{1-2/\alpha})\right )  \frac{d\omega}{2\pi} \notag \\
\end{eqnarray}
Rescaling the integration variable gives
\begin{eqnarray}
f(T,N) & = & \int_{-\infty}^{\infty} exp\left (i  \omega \left(T + \frac{\alpha}{1-\alpha}  t_{0} N\right) +\alpha \left (i \omega t_{0}\right)^{\alpha} \Gamma(-\alpha)N\right )  \frac{d\omega}{2\pi} \notag \\
\end{eqnarray}
which is the Levy-stable distribution with stability parameter $\alpha$. Defining the spatial variable $r \equiv N \delta r$, it is easy to see that this distribution satisfies the \emph{fractional diffusion temporal equation}
\begin{eqnarray}
\frac{\partial }{\partial r}f\left( T;r/\delta r\right)
= \left[ \frac{\alpha t_{0}}{(1-\alpha) \delta r}\frac{\partial }{\partial T}+\frac{{\alpha}t_{0}^{\alpha }}{ \delta r}%
\Gamma \left( -\alpha \right) \frac{\partial
^{\alpha }}{\partial T^{\alpha }}\right] f_{\alpha }\left( T;r/\delta
r\right)  \;.
\label{17}
\end{eqnarray}
Notice that, with time and space variables interchanged, the FFPE exhibits a structure analogous 
to the fractional Fokker-Plank equation for anomalous spatial diffusion that follows from the 
continuous time random walk model with a power law {\it ansatz} for the waiting times \cite{metzler}. 

\section{Conclusions}
\label{conclusion}

When considering diffusion processes from the viewpoint of a temporal formulation -
dual to the classical spatial description - a Fokker-Plank equation (FPE) description is 
found to be equally valid for temporal diffusion. In the latter case
the FPE exhibits a solution for the temporal distribution function showing Gaussian 
behavior \cite{boon-grosfils-lutsko} similar to the Gaussian solution of the classical
diffusion equation, but with time and space reversed. However when, as in most real systems, 
the diffusive medium is inhomogeneous, this classical description is modified because
the dynamics, and consequently the corresponding distribution function, may depend 
on the local concentration variations in time and and space and on the distribution
of time delays in the diffusive process. We considered both types of dependences.
(i) Starting from the classical random walk model, we generalized Einstein's master
equation by including a functional concentration dependence in the jump probability
wherefrom a temporal nonlinear Fokker-Plank equation is obtained and solved to yield
the temporal distribution function evolving from Gaussian shape to finite support when
the nonlinear exponent increases. (ii) On the other hand using a power law waiting time
probability distribution we obtain a fractional temporal Fokker-Plank equation similar to
the usual fractional Fokker-Plank equation \cite{metzler} with space and time interchanged.
These results should provide insight for the elucidation of the mechanisms of
temporal diffusion processes.


\appendix

\section{Expansion of master equation}

\label{appendixA}

We first consider the expansion of $F_{j}^{(\nu)}\left[ f\right] $ (for simplicity 
in the notation we shall omit the upper index ${(\nu)}$ which will be reintroduced 
when necessary):%
\begin{align}
F_{j}\left[ f\right] & =\left\{ F_{j}\left( x;y_{1}...,y_{N}\right) \right\}
_{f} 
 -\delta t\left\{ j\frac{\partial F_{j}\left( x;y_{1}...,y_{N}\right) }{%
\partial x}+\sum_{l=1}^{N}l\frac{\partial F_{j}\left(
x;y_{1}...,y_{N}\right) }{\partial y_{l}}\right\} _{f}\left(\frac{\partial f\left(
r,t\right) }{\partial t} \right) +...  \,, 
\label{norm_cond}
\end{align}
where the notation $\left\{ ...\right\} _{f}$ means that all the variables $%
x,y_{1},...$ are to be set equal to the $f\left( r,t\right) $'s in the r.h.s. of Eq.\Eq{Fnu}.
Using this expansion, the generalized master equation \Eq{a4} becomes%
\begin{align}
&\delta r\frac{\partial f\left( r,t\right) }{\partial r}+\frac{1}{2}\left(
\delta r\right) ^{2}\frac{\partial^{2}f\left( r,t\right) }{\partial r^{2}}%
+... =-\delta t\sum_{j=1}^{N}jp_{j}\left\{ F_{j}\left(
x;y_{1}...,y_{N}\right) \right\} _{f}\frac{\partial f\left( r,t\right) }{%
\partial t} \notag\\
& +\frac{1}{2}\left( \delta t\right) ^{2}\sum_{j=1}^{N}jp_{j}\left\{
F_{j}\left( x;y_{1}...,y_{N}\right) \right\} _{f}\frac{\partial^{2}f\left(
r,t\right) }{\partial t^{2}} \notag \\
& +\left( \delta t\right) ^{2}\sum_{j=1}^{N}jp_{j}\left\{ j\frac{\partial
F_{j}\left( x;y_{1}...,y_{N}\right) }{\partial x}+\sum_{l=1}^{N}l\frac{%
\partial F_{j}\left( x;y_{1}...,y_{N}\right) }{\partial y_{l}}\right\}
_{f}\left( \frac{\partial f\left( r,t\right) }{\partial t}\right) ^{2} 
+...  \,.
\end{align}

We now perform a multiscale expansion with
\beq{rescale}
\frac{\partial}{\partial r} \rightarrow \epsilon \frac{\partial}{\partial {r_1}} \,+\,\epsilon^2 \frac{\partial}{\partial {r_2}}
\;\;\; ;\;\;\; \frac{\partial}{\partial t} \rightarrow \epsilon \frac{\partial}{\partial {t_1}} \,+\,\epsilon^2 \frac{\partial}{\partial {t_2}} \,,
\eeq 
and $f\,=\,f^{(0)}\,+\,\epsilon f^{(1)}\,+\,\,{\cal O}(\epsilon^2)$, where
$f^{(0)}$ is the distribution function in the absence of dispersion. 
To first order, we obtain %
\begin{equation}
{\cal O}(\epsilon^{1})  :  \;\;\;\;\;\; \delta r\frac{\partial f^{(0)}\left( r,t\right) }{\partial r_{1}}=-\delta t \sum_{j=1}^{N}jp_{j}\left\{ F_{j}\left( x;y_{1}...,y_{N}\right) \right\}_{f^{(0)}}\frac{\partial f^{(0)}\left( r,t\right) }{\partial t_{1}}  \,,
\label{ms1st}
\end{equation}%
and to second order%
\begin{align}
&{\cal O}(\epsilon^{2})  :  \;\;\;\;\;\; \delta r\frac{\partial f^{(1)}\left( r,t\right) }{\partial r_{1}}+\delta r\frac{\partial f^{(0)}\left( r,t\right) }{\partial r_{2}}+\frac{1}{2}\left(
\delta r\right) ^{2}\frac{\partial ^{2}f^{(0)}\left( r,t\right) }{\partial r_{1}^{2}} \notag \\
& =-\delta t\sum_{j=1}^{N} j p_{j}\left\{ \frac{\partial F_{j}\left(
x;y_{1}...,y_{N}\right) }{\partial x}+\sum_{l=1}^{N}  \frac{\partial
F_{j}\left( x;y_{1}...,y_{N}\right) }{\partial y_{l}}\right\}_{f^{(0)}}%
 f^{(1)}\left( r,t\right) \frac{\partial f^{(0)}\left( r,t\right) }{\partial t_{1}} \notag \\
& -\delta t\sum_{j=1}^{N}jp_{j}\left\{ F_{j}\left( x;y_{1}...,y_{N}\right)
\right\} _{f} \left(\frac{\partial f^{(1)}\left( r,t\right) }{\partial t_{1}}%
+\frac{\partial f^{(0)}\left( r,t\right) }{\partial t_{2}} \right) \notag \\
& +\frac{1}{2}\left( \delta t\right) ^{2}\sum_{j=1}^{N}j^{2}p_{j}\left\{
F_{j}\left( x;y_{1}...,y_{N}\right) \right\} _{f}\frac{\partial
^{2}f^{(0)}\left( r,t\right) }{\partial t_{1}^{2}}  \notag \\
& +\left( \delta t\right) ^{2}\sum_{j=1}^{N}jp_{j}\left\{ j\frac{\partial
F_{j}\left( x;y_{1}...,y_{N}\right) }{\partial x}+\sum_{l=1}^{N}l\frac{%
\partial F_{j}\left( x;y_{1}...,y_{N}\right) }{\partial y_{l}}\right\}
_{f^{(0)}}\left( \frac{\partial f^{(0)}\left( r,t\right) }{\partial t_{1}}%
\right) ^{2}  \,.
\label{ms2nd}
\end{align}%
From the normalization condition (with \Eq{norm_cond} where
$\frac{\partial f\left( r,t\right) }{\partial t}$ is unconstrained) we have%
\begin{align}
&1 = \sum_{j=1}^{N}p_{j}\left\{ F_{j}\left( x;y_{1}...,y_{N}\right) \right\}_{f^{(0)}} \,,\notag \\
&0 = \sum_{j=1}^{N}p_{j}\left\{ j\frac{\partial F_{j}\left(
x;y_{1}...,y_{N}\right) }{\partial x}+\sum_{l=1}^{N}l\frac{\partial
F_{j}\left( x;y_{1}...,y_{N}\right) }{\partial y_{l}}\right\} _{f^{(0)}} \,.
\label{norm1}
\end{align}%
It is easy, for instance, to check that these relations are indeed verified in the case of
the power law (\ref{nu_norm}). Differentiating (\ref{ms1st}) with respect to $r_1$ and reinserting (\ref{ms1st}) in the result, we obtain
\begin{align}
&\left( \delta r\right) ^{2}\frac{\partial ^{2}f^{(0)}\left( r,t\right) }{%
\partial r_{1}^{2}} =-\delta r\delta t\sum_{j=1}^{N}jp_{j}\left\{
F_{j}\left( x;y_{1}...,y_{N}\right) \right\} _{f^{(0)}}\frac{\partial
^{2}f^{(0)}\left( r,t\right) }{\partial t_{1}\partial r_{1}} \notag \\
& =\left( \delta t\right) ^{2}\sum_{j=1}^{N}jp_{j}\left\{ F_{j}\left(
x;y_{1}...,y_{N}\right) \right\} _{f^{(0)}}\frac{\partial }{\partial t_{1}}%
\left\{ \sum_{l=1}^{N}lp_{l}F_{l}\left( x;y_{1}...,y_{N}\right) \right\}
_{f^{(0)}}\frac{\partial f^{(0)}\left( r,t\right) }{\partial t_{1}}  \notag \\
& =\left( \delta t\right) ^{2}\sum_{j=1}^{N}jp_{j}\left\{ F_{j}\left(
x;y_{1}...,y_{N}\right) \right\} _{f^{(0)}} \times \notag \\
&\;\;\;\;\;\;\;\;\; \times \sum_{l=1}^{N}lp_{l} \left\{\frac{\partial F_{l}\left( x;y_{1}...,y_{N}\right) }
{\partial x}+\sum_{k=1}^{N}\frac{\partial F_{l}\left(
x;y_{1}...,y_{N}\right) }{\partial y_{k}}\right\} _{f^{(0)}} 
 \left( \frac{\partial f^{(0)}\left( r,t\right) }{\partial t_{1}}\right) ^{2}  \notag \\
& +\left( \delta t\right) ^{2}\left( \sum_{j=1}^{N}jp_{j}\left\{ F_{j}\left(
x;y_{1}...,y_{N}\right) \right\} _{f^{(0)}}\right) ^{2}\frac{\partial
^{2}f^{(0)}\left( r,t\right) }{\partial t_{1}^{2}}  \,.
\end{align}%
Using this result in (\ref{ms2nd}) gives
\begin{align}
& \delta r\frac{\partial f^{(1)}\left( r,t\right) }{\partial r_{1}}+\delta r%
\frac{\partial f^{(0)}\left( r,t\right) }{\partial r_{2}}  \label{ms2} \notag \\
& =-\delta t\sum_{j=1}^{N}jp_{j}\left\{ F_{j}\left( x;y_{1}...,y_{N}\right)
\right\} _{f^{(0)}}\ \left( \frac{\partial f^{(1)}\left( r,t\right) }{\partial t_{1}}%
+ \frac{\partial f^{(0)}\left( r,t\right) }{\partial t_{2}} \right) \notag \\
&-\delta t\sum_{j=1}^{N} j p_{j}\left\{ \frac{\partial F_{j}\left(
x;y_{1}...,y_{N}\right) }{\partial x}+\sum_{l=1}^{N}  \frac{\partial
F_{j}\left( x;y_{1}...,y_{N}\right) }{\partial y_{l}}\right\}_{f^{(0)}}%
 f^{(1)}\left( r,t\right) \frac{\partial f^{(0)}\left( r,t\right) }{\partial t_{1}} \notag \\
& +\frac{1}{2}\left( \delta t\right) ^{2}\left(
\sum_{j=1}^{N}j^{2}p_{j}\left\{ F_{j}\left( x;y_{1}...,y_{N}\right) \right\}
_{f^{(0)}}-\left( \sum_{j=1}^{N}jp_{j}\left\{ F_{j}\left(
x;y_{1}...,y_{N}\right) \right\} _{f^{(0)}}\right) ^{2}\right) \frac{%
\partial ^{2}f^{(0)}\left( r,t\right) }{\partial t_{1}^{2}}  \notag \\
& +\left( \delta t\right) ^{2}\left( 
\begin{array}{c}
\sum_{j=1}^{N}jp_{j}\left\{ j\frac{\partial F_{j}\left(
x;y_{1}...,y_{N}\right) }{\partial x}+\sum_{l=1}^{N}l\frac{\partial
F_{j}\left( x;y_{1}...,y_{N}\right) }{\partial y_{l}}\right\} _{f^{(0)}} \\ 
-\sum_{j=1}^{N}jp_{j}\left\{ F_{j}\left( x;y_{1}...,y_{N}\right) \right\}
_{f^{(0)}} \sum_{l=1}^{N}lp_{l}\left\{ \frac{\partial F_{l}\left(
x;y_{1}...,y_{N}\right) }{\partial x}+\sum_{k=1}^{N}%
\frac{\partial F_{l}\left( x;y_{1}...,y_{N}\right) }{\partial y_{k}}\right\}_{f^{(0)}}%
\end{array} \right) \times \notag \\
& \;\;\;\;\;\;\;\;\;\;\; \times \left( \frac{\partial f^{(0)}\left( r,t\right) }{\partial t_{1}}%
\right) ^{2}  \,,
\end{align}
After recombining first and second order terms, resummation  yields Eq.\Eq{GFPE1}.

\newpage

\section{General confluent equation}
\label{appendixB}

\begin{equation}
\frac{d^{2}w}{dx^{2}}+x\frac{dw}{dx}+\left( 2\nu - 1 \right) w=0 \,,
\label{eq_w}
\end{equation}%

From  AS 13.1.35 in \cite{abramow}, the general confluent equation is 
\beqa
0 &=&  w^{\prime \prime }+\left( \frac{2a}{y}+2f^{\prime }(y)+\frac{bh^{\prime
}(y)}{h(y)}-h^{\prime }(y)-\frac{h^{\prime \prime }(y)}{h^{\prime }(y)}%
\right) w^{\prime } \nm
&+& \left( \frac{bh^{\prime }(y)}{h(y)}-h^{\prime }(y)-\frac{h^{\prime
\prime }(y)}{h^{\prime }(y)}\right) \left( \frac{a}{y}+f^{\prime }(y)\right) w \nm
&+& \left( \frac{a(a-1)}{y^{2}}+\frac{2af^{\prime }(y)}{y}+f^{\prime \prime
}(y)+f^{\prime 2}(y)-\frac{c h^{\prime 2}(y)}{h(y)}\right) w  \,. 
\eeqa%
This equation and Eq.\Eq{eq_w} match provided 
\begin{eqnarray}
a = 0 \;\;\;;\;\;\;\;
b = \frac{1}{2}  \;\;\;;\;\;\;\;
c = 1 - {\nu}  \;\;\;;\;\;\;\;
f(y) = h(y)=\frac{y^{2}}{2}  \,, \notag
\end{eqnarray}%
and the general solution then is 
\begin{equation}
w\left( y\right) =A\exp \left( -\frac{y^{2}}{2}\right) M\left( {1 - \nu},\frac{1}{2},\frac{y^{2}}{2}\right) 
+B\exp \left( -\frac{y^{2}}{2}\right) U\left(1 - {\nu},\frac{1}{2},\frac{y^{2}}{2}\right)\,,  
\end{equation}%
or, using AS 13.1.3 in \cite{abramow}, with $y \equiv x$,
\begin{equation}
w\left( x\right) = {A}\,\left[M\left( {1 - {\nu}},\frac{1}{2},\frac{x^{2}}{2}\right)%
+x \,\frac{B}{A}\, M\left( \frac{3}{2}- {\nu},\frac{3}{2},\frac{x^{2}}{2}\right) \right]%
\exp \left( -\frac{x^{2}}{2}\right) \,. 
\end{equation}%
Note that $w\left( 0\right)  = A $ and $w^{\prime }\left( 0\right)=~B $.

\end{document}